# Title: Single-molecule study on the orientation of coiled-coil proteins attached to gold nanostructures

*Jae-Byum Chang[+], Yong Ho Kim[+], Evan Thompson, Young Hyun No, Nam Hyeong Kim, Jose Arrieta, Vitor R. Manfrinato, Amy E. Keating\*, Karl K. Berggren\* ([+]equally contributed)*

Jae-Byum Chang
Department of Materials Science and Engineering
Massachusetts Institute of Technology
Cambridge MA 02139, USA

Prof. Yong Ho Kim, Young Hyun No,
SKKU Advanced Institute of Nanotechnology (SAINT)
Sungkyunkwan University
Suwon 440-746, Republic of Korea

Prof. Yong Ho Kim, Nam Hyeong Kim
Department of Chemistry
Sungkyunkwan University
Suwon 440-746, Republic of Korea

Evan Thompson, Prof. Amy E. Keating
Department of Biology
Massachusetts Institute of Technology
Cambridge, MA 02139, USA
E-mail: keating@mit.edu

Prof. Amy E. Keating
Department of Biological Engineering
Massachusetts Institute of Technology
Cambridge, MA 02139, USA

Jose Arrieta, Vitor R. Manfrinato, Prof. Karl K. Berggren
Department of Electrical Engineering and Computer Science
Massachusetts Institute of Technology
Cambridge, MA 02139, USA
E-mail: berggren@mit.edu

**Abstract**

Methods for patterning biomolecules on a substrate at the single molecule level have been studied as a route to sensors with single-molecular sensitivity or as a way to probe biological phenomena at the single-molecule level[1–3]. However, the arrangement and orientation of single biomolecules on substrates has been less investigated. Here, we examined the arrangement and orientation of two rod-like coiled-coil proteins, cortexillin and tropomyosin, around patterned gold nanostructures. The high aspect ratio of the coiled coils made it possible to study their orientations and to pursue a strategy of protein orientation via two-point attachment. The proteins were anchored to the surfaces using thiol groups, and the number of cysteine residues in tropomyosin was varied to test how this variation affected the structure and arrangement of the surface-attached proteins. Molecular dynamics studies were used to interpret the observed positional distributions. Based on initial studies of protein attachment to gold post structures, two 31-nm-long tropomyosin molecules were aligned between the two sidewalls of a trench with a width of 68 nm. Because the approach presented in this study uses one of twenty natural amino acids, this method provides a convenient way to pattern biomolecules on substrates using standard chemistry.

## 1. Introduction

Patterning biomolecules at the nanoscale on solid substrates is of interest for applications including studying biological phenomena at the single-molecule level[4–6], building sensors at the single-molecule level[2,7–9], and patterning sub-5-nm inorganic particles at the single-particle level[10,11]. Patterning individual DNA or protein molecules on substrates enables the study of the kinetics of biological events at the single-molecule level, rather than only at the level of ensembles[5,6]. Single-biomolecule patterning also enables single-DNA-strand sequencing, which can accelerate and increase the fidelity of sequencing reactions[2,8]. In addition, the functionalities of biomolecules can be harnessed to pattern inorganic materials at the single-molecule level, by patterning single proteins or DNA strands on substrates and then attaching inorganic nanoparticles, such as quantum dots, to the patterned biomolecules[10,11].

The general approach to patterning proteins at the single-molecule level is to attach biomolecules of interest to lithographically defined nanostructures through unstructured chemical linkers, e.g. using a streptavidin-biotin-linker complex[12–18], or by adsorption[19]. Such attachment methods are not expected to impart a prefered orientation to molecules attached to surfaces. To our knowledge, there have been no prior studies of how proteins attached to nanostructures are assembled and oriented at the single-molecule level.

In this work, we studied the orientations of three coiled-coil protein variants of two lengths attached to patterned nanostructures. The three coiled-coil proteins were attached to gold nanostructures through cysteine-gold interactions, and their orientations were studied. We investigated how the structure and orientation of surface-attached tropomyosin varied with the number of possible cysteine-gold bonds per molecule. Based on these studies, we used a gold trench to position a 31-nm-long coiled-coil protein. Images of gold particles attached to the coiled-coil proteins were consistent with a novel trench-spanning architecture.

## 2. Results and Discussion
### 2.1. Attaching coiled-coil proteins to gold nanostructures

Our study used two coiled-coil proteins to examine how biomolecules attached to nanostructures can be assembled and organized. A coiled-coil is a common structural motif formed by approximately 2-3% of amino acids in proteins[20]. Coiled coils consist of 2-5 α-helices supercoiled around one another[21]. Proteins containing coiled-coil domains are involved in a diverse range of biological functions[22,23]. Structural proteins that contain coiled coils are an important part of the cytoskeleton. Among various coiled-coil proteins, cortexillin I and tropomyosin were used in this study. Cortexillin I is a member of the α-actin/spectrin superfamily that binds directly to actin and modulates the organization and function of the actin-myosin system. It consists of a large globular domain, a rod-like coiled-coil domain, and a relatively flexible tail domain[24]. In this study, we used only the coiled-coil domain, which spans residues 272 to 353 (denoted cortexillin I$_{272-353}$). Tropomyosin is a coiled-coil protein that wraps around actin filaments and regulates muscle contraction[25]. The average end-to-end distances of cortexillin I$_{272-353}$ and tropomyosin, as determined by a molecular dynatmics (MD) simulation, were 15 nm and 31 nm, respectively (see supplementary Figure S1). The dimensions of these proteins match length scales that can be patterned by e-beam lithography, and coiled coils of this size that are labeled with gold nanoparticles as described below are readily imaged by scanning electron microscopy (SEM).

To attach the coiled-coil proteins to gold nanostructures and facilitate labeling for visualization, a cysteine residue was added to the N-terminus of cortexillin I$_{272-352}$ and the C-terminus of tropomyosin, as shown in **Figure 1**(a). In contrast to the coiled-coil region of native cortexillin I$_{272-352}$, which does not contain any cysteine residues, tropomyosin has an internal cysteine residue at position 190. We prepared a mutant protein, tropomyosin C190A, in which the C-terminus included a cysteine residue but the internal cysteine was replaced with alanine. FLAG-tag and His$_6$ tag were added to the C-terminus of cortexillin I$_{272-352}$ and the N-terminus of native and C190A tropomyosin, as shown in **Figure 1**(a). The FLAG motif was used for recognition by an anti-FLAG antibody, and the His-tag motif was used to attach Ni-nitrilotriacetic acid(NTA)-coated gold nanoparticles to the proteins. Various gold nanostructures were fabricated on silicon substrates by using conventional electron-beam lithography and lift-off process[26].

We first used fluorescence microscopy to determine whether the coiled-coil proteins could be attached to gold nanostructures by cysteine-gold interaction. Gold nanodot arrays with various diameters and pitches were fabricated on a silicon substrate and incubated with cortexillin I$_{272-352}$ (see Methods for details). The diameters of the gold nanodots were 40 nm, 60 nm, and 80 nm, and the pitches of the gold nanodot arrays were 100 nm, 200 nm, and 250 nm, as shown in **Figure 2**(a,b). After the incubation of the substrate with cortexillin, the substrate was then incubated with a primary antibody against FLAG, and then a FITC-tagged secondary antibody. The immunostained substrate was imaged by using an fluorescence microscope. In Figure 2(c), bright yellow represents a strong fluorescence signal from the patterns, and dark purple represents a weak signal. The strong fluorescence signal around the gold nanodot arrays indicates that the cortexillin was attached selectively to the gold patterns.

**2.1. Arrangement of coiled-coil proteins attached to gold nanostructures**

To understand the orientations of the proteins around the gold nanopatterns at the single-molecule level, proteins attached to nanostructures were tagged with gold nanoparticles and imaged by SEM. Gold nanodot arrays with a diameter of 100 nm were fabricated on a silicon substrate and incubated with cortexillin I$_{272-352}$. The substrate was then incubated with Ni-NTA-coated gold nanoparticles with a diameter of 5 nm. As shown in **Figure 3**(a), gold nanoparticles were observed around the patterned gold nanostructures, but not in other areas of the substrate, consistent with observations made using fluorescence imaging. To further understand the orientations of cortexillin attached to the gold nanodots, we measured the size of the gaps between the gold nanoparticles and gold nanodots. As shown in Figure 3(b), the average gap size was 9.2 nm, which is 61% of the end-to-end distance of the cortexillin I$_{272-352}$, measured by a MD simulation. To study the effect of the denaturation of proteins, we incubated cortexillin I$_{272-352}$ with the patterned substrate at 80°C. When cortexillin I$_{272-352}$ was incubated at 80°C, the average gap size decreased to 5 nm. A similar trend was observed when cortexillin I$_{272-352}$ was incubated with the patterned substrate in 8 M urea.

The longer tropomyosin protein was studied in the same way. When native tropomyosin was incubated with the same patterned substrate and Ni-NTA-coated gold nanoparticles, the gold nanoparticles were assembled around the gold nanodots with an average gap size of 10 nm, (**Figure 4**(a,b)). However, when tropomyosin C190A, in which the internal cysteine was replaced with alanine, was incubated with the patterned substrate, the gold nanoparticles were assembled around the patterned gold nanodots with a larger gap (Figure 4c,d). The average gap size was 21 nm, which is 68% of the end-to-end distance of the C190A tropomyosin measured by a MD simulation. This ratio for C190A tropomyosin was simliar to the ratio of the measured gap size to the simulated end-to-end length of cortexillin I$_{272-353}$, which is 60%. To understand the small gap sizes between the gold nanoparticles and gold nanostructures for both cortexillin I$_{272-352}$ and C190A tropomyosin, MD simulations were used to study the conformations of cortexillin I$_{272-352}$ and tropomyosin in solution. When the N-termini of 120 simulated cortexillin I$_{272-352}$ and tropomyosin structures were superimposed, both proteins occupied cone-shaped volumes due to their flexible backbones, as shown in **Figure 5**(a,b). We calculated the distances between the N-terminus of the simulated tropomyosin to a gold nanosphere with a diameter of 100 nm, if the C-terminus of the tropomyosin was attached to the gold nanodot. The average distance was 19 nm, which is consistent with the experimental result shown in Figure 4(d). This result suggests that the smaller gap sizes between the gold nanoparticles and gold nanodots can be attributed to the flexible backbones of the proteins.

**2.3. Coiled-coil protein templated by a gold trench**
Based on the experiments and simulation results above, we attempted control of the orientation of a coiled-coil protein by using a gold trench. As shown in **Figure 6**, a gold trench of continuously varying width, from 40 to 90 nm, was fabricated and incubated with tropomyosin C190A and gold nanoparticles. The trench was designed so that, at appropriate widths, two tropomyosin coiled coils could span it and interact with a shared gold nanoparticle. As shown in Figure 6(a), the gold nanoparticles tended to be aligned at the center of the trench, when the trench width was around 68 nm. When the trench width was narrower or wider than 68 nm, the gold nanoparticles were not aligned at the center. Instead, they tended to be randomly scattered inside the trench (see Supplementary Figure S2,3 for details). The fact that 68 nm is close to double the average end-to-end distance of the C190A tropomyosin suggests that the tropomyosin C190A molecules were aligned perpendicularly to the trench and the gold nanoparticles were trapped by the two C190A tropomyosin molecules attached to the two sides of the trench.

**3. Conclusion**

We demonstrated the attachment of coiled-coil proteins to gold nanostructures using cysteine-gold interaction, and visualized the attached proteins using SEM imaging of gold nanoparticles bound to the terminal His-tag motifs of the proteins. We also studied the change of observed patterns depending on the number of possible cysteine interactions with the gold nanostructures. The likely conformations of the proteins were studied by using MD simulations and matched with the experimental results. Based on this study, we implemented a scheme to control the orientation of 31-nm-long coiled-coil proteins by using a gold trench and observed distributions of protein-attached gold nanoparticles that were consistent with this scheme.

We report here a proof-of-concept study, but we expect that the process used in this study can be developed as the basis for patterning a variety of proteins with different functions. Coiled coils have special properties that make them suitable for orientation on lithographically patterned surfaces, but coiled coils can be readily linked to other molecules such as photoactive protein or enzymes. Dynamic control of modified surfaces could be achieved by releasing the attached proteins by controlling the pH or temperature of the substrates,[27] or by selectively increasing the temperature of the nanostructures by using light[28]. Our understanding of surface-controlled protein orientation could be further improved by considering the effect of the surface and nanostructures on the substrate in MD simulation. When improved, the approach demonstrated in this study could be used to develop a protein array for studying biological phenomena at the single-molecule level or developing a protein sensor with single-molecule-level sensitivity.

More precise single-molecule patterning might be achieved by patterning nanostructures with different chemical affinities for proteins. For example, a pair comprised of a gold nanodot and an NHS ester-functionalized silicon nanodot[29] could be used to attach a single protein in a controlled orientation through cysteine-gold interaction and lysine-NHS ester conjugation[30]. This approach could be further improved by taking advantage of protein modification strategies and state-of-the-art nanofabrication techniques that have already been developed[31]. Once developed, such precise single-molecule patterning techniques could be used to mimic the arrangements of proteins inside cells in order to study biological phenomena on a substrate, to replicate the arrangements of biosynthetic enzymes inside cells in order to produce useful biomolecules, or to pattern inorganic or other organic molecules by using the precisely patterned biomolecules as templates.

## 4. Experimental Section

*Gold structure fabrication*: A silicon substrate was spin-coated with a thin layer of a positive-tone electron-beam lithography resist poly(methyl methacrylate)(MicroChem 950PMMA) with a thickness of 60 nm. The spin-coated resist film was exposed with a Raith 150 electron-beam lithography system operating at 30 kV acceleration voltage (Figure S4a). The exposed resist film was developed in 3:1 isopropyl alcohol (IPA):methyl isobuthyl ketone (MIBK) at -15˚C for 30 sec (Figure S4b)[32]. 3-nm titanium film and 6-nm gold film were deposited onto the sample by physical vapor deposition (Figure S4c). Then, the metal films on top of the resist film were removed by a standard lift-off process (Figure S4d)[26]. The lift-off process was carried out in 60˚C n-methylpyrrolidone (NMP) for 20 min. The sample was rinsed with acetone, and then sonicated in acetone at room temperature for 1 min to completely remove any residual metal films.

*Protein preparation*: Synthetic genes for the native and C190A mutant human tropomyosin alpha-1 genes were synthesized (Bioneer Inc.) and amplified by PCR. PCR was performed under the following conditions: 98˚C for 5 min, 30 cycles of 98˚C for 1 min (denaturation), 58˚C for 50 sec (annealing), 72˚C for 1 min (extension), followed by additional extension for 72˚C for 10 min and finally incubation at 4˚C. The amplified PCR product was cleaved using DNA restriction enzymes Nde1 (5' cut) and EcoR1 (3' cut) and purified using QIAquick PCR Purification Kit (QIAGEN). The purified native and C190A mutant genes contained a N-terminal His$_6$-tag and thrombin cleavage site and were ligated into pET-43.1 a(+) vector (See Supplementary Information for the sequence). DNA sequencing of the PCR products and plasmids was done at GENEWIZ Inc. The construct was transformed into BL21 (DE3) *E.coli*. The cells were grown in Difco$^{tm}$ LB (Luria-Bertani) broth Miller media containing 1 mM ampicillin at 37˚C for 3-4 hours until OD$_{600}$ reached 0.8~1. Isopropyl b-D-1-thiogalactopyranoside (IPTG) in total concentration of 1 mM was added in the media and the cells were cultured for another 6 hours. After the incubation at 37˚C for 6 hours, the cells were centrifuged at 5000 rpm for 15 min and lysed by sonication in 20 mM Tris, 500 mM NaCl, 5 mM imidazole, 1 mM DTT pH 8.0 buffer (binding buffer). The lysate was cleared by centrifugation. The supernatant was loaded onto a Ni-NTA affinity column and eluted in 20 mM Tris, 500 mM NaCl, 300 mM imidazole pH 8.0 buffer (elution buffer). The size of proteins was confirmed by sodium dodecyl sulfate-polyacrylamide gel electrophoresis (SDS-PAGE) (See Figure S5). We also checked the purity of proteins by using reverse-phase high-performance liquid chromatography (RP-HPLC) on a C3 column (ZORBAX 300SB-C3, 9.4 mm by 250 mm, Agilent Technologies, USA). Proteins were eluted with 90% acetonitrile at a flow rate of 1 ml/min using a 20–60% linear gradient of elution buffer over 40 min. The purified proteins were lyophilized and stored at -20˚C.

*PEGylation*: Patterned substrate was cleaned and hydroxylated by brief (~10 sec) UV-ozone treatment (Figure S4e). Methoxyl silane polyethylene glycol (PEG) with a molecular weight of 5000 (Nanocs PG1-SL-5k) was dissolved in 20 ml of anhydrous toluene. The UV-ozone treated substrate was placed in a glass vial and 2 ml of the PEG solution and 2 µL of acetic acid were added to the vial. The vial was incubated overnight at room temperature. After the incubation, the substrate was washed first in anhydrous toluene, second in ethanol, and lastly in acetone (Figure S4f).

*Protein treatment*: Protein solutions were prepared by dissolving lyophilized protein powder in 20 mM Tris, 100 mM NaCl, 1 mM tris(2-carboxyethyl)phosphine (TCEP), 0.1% TWEEN 20 (working buffer). The concentrations of the protein solutions were measured by measuring the UV absorbances of the solutions (NanoDrop 1000, NanoDrop, USA) and adjusted to 1 µM. The pegylated substrate was placed in a clean glass vial and 2 ml of the protein solution was added to the glass vial. The vial was incubated at room temperature overnight. After incubation, the substrate was washed in working buffer (Figure S4g).

*Fluorescence antibody labeling and imaging*: The protein-treated substrate was incubated with rabbit anti-FLAG antibody (Sigma F7425-2MG) in tris-buffered saline plus 0.1% TWEEN20 (staining buffer) at a concentration of 1 µg/ml for 1 hour at room temperature, and then washed with staining buffer. The substrate was incubated with fluorescein isothiocyante (FITC)-conjugated anti-rabbit antibody in staining buffer at a concentration of 1 µg/ml for 1 hour at room temperature, and then washed in staining buffer (Figure S4i). The stained substrate was imaged on a Zeiss AxioPlan2 upright microscope.

*Gold particle labeling and imaging*: Ni-nitrilotriacetic acid (NTA) conjugated gold nanoparticles with a diameter of 5 nm (Nanoprobes, USA) were dissolved in 20 mM Tris, 100 mM NaCl at a concentration of 50 nM. Protein-treated substrate was placed in a glass vial and 2 ml of the gold nanoparticle solution was added to the vial. After 1 hour incubation at room temperature, the substrate was washed in 20 mM Tris, 100 mM NaCl and dried in a stream of nitrogen (Figure S4h). The SEM images were acquired via a Raith 150 electron-beam lithography system operating at 10 kV acceleration voltage.

*Molecular dynamics (MD) simulation*: MD simulations were carried out with the NAMD package using the CHARMM27 force field to simulate the structure and flexibility of coiled-coil proteins. Cortexillin and tropomyosin structures were prepared by using the crystallized structures of the two proteins reported in protein data bank (PDB). We used PDB structure 1D7M for cortexillin and 1C1G for tropomyosin[33,34]. We ran 30-ns trajectories using a generalized born implicit solvent (GBIS) model[35,36]. GBIS parameters were as follows: ion concentration=0.3 M, cutoff=14 Å, and alphacutoff=12 Å. After structural relaxation for 20 ps at 300 K, equilibrium dynamics simulations were carried out for an NVT ensemble at 300K for 30 ns. For analysis of coiled-coil conformers, frames were collected every 250 ps during the 30 ns simulations. To study the conformations of the protein when its C-terminus is attached to a gold nanostructure, the collected conformers were superimposed at 1-10 N-terminal residues in cortexillin and 1-28 C-terminal residues in tropomyosin, respectively, which correspond to 10% of total protein residue numbers[25]. End-to-end distances of both proteins were measured from N-terminus to C-terminus, using the $C_\alpha$ atoms of each terminus residues, to generate the distribution in Figure 5(c,d).

**Acknowledgements**

J.-B.C and Y.H.K. contributed equally to this work. This study was supported by International Iberian Nanotechnology Laboratory (INL) (K.K.B. and A.E.K.) and National Science Foundation (NSF) (MCB-0950233) (A.E.K.). This research was also supported by Basic Science Research Program through the National Research Foundation of Korea (NRF) funded by the Ministry of Education (NRF-2014R1A1A2055647). This work was supported by the National Research Foundation of Korea (NRF) grant funded by the Korea government (MSIP) (No. 2009-0083540). J.-B.C was funded by a scholarship from Samsung Scholarship Foundation. The authors thank to Mark Mondol and Jim Daley for their technical supports. The Research Laboratory of Electronics Scanning-Electron-Beam Lithography Facility at MIT provided facilities for this work.

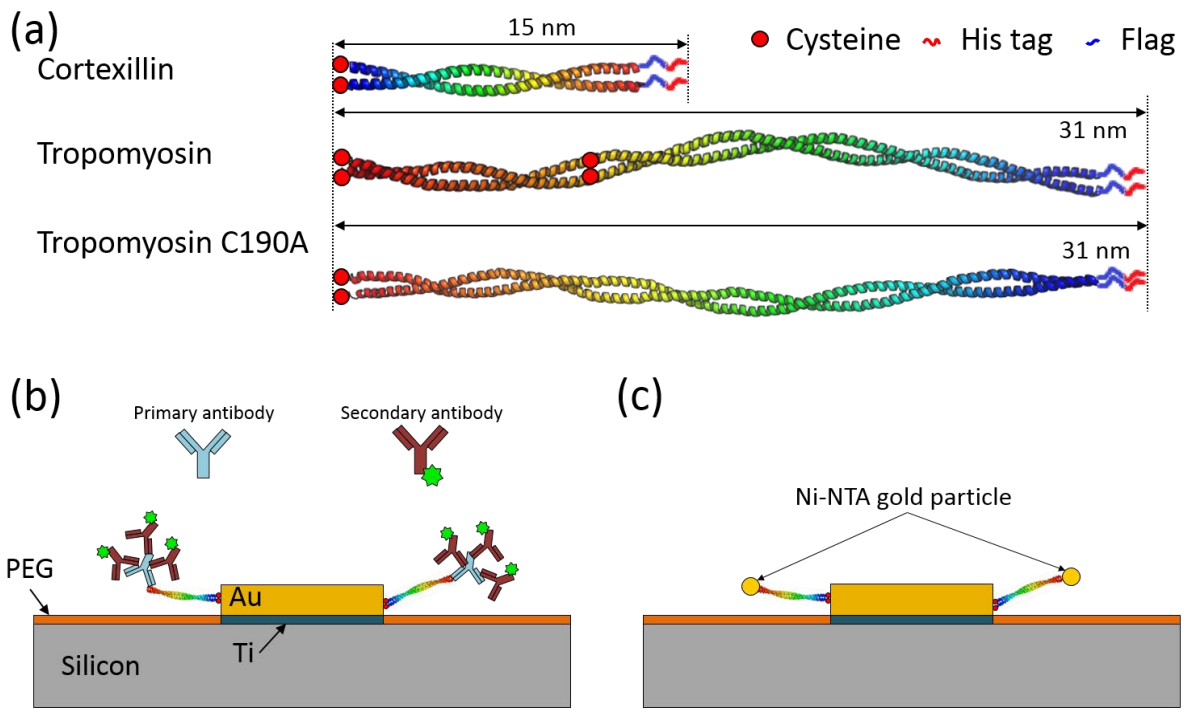

**Figure 1**. Proteins, gold nanostructures, and indirect protein visualization methods used in this study. (a) Diagram of the three proteins used in this study. The average end-to-end distance was determined by MD simulation. The 3D structures of the three proteins were rendered with rainbow spectrum (blue: N-terminus, red: C-terminus). (b) Diagram showing the immunostaining protocols used to detect proteins anchored to gold nanostructure surfaces. Primary antibody was rabbit anti-Flag antibody and secondary antibody was fluorescein isothiocyante (FITC, green star in the diagram)-conjugated anti-rabbit IgG antibody. (c) Diagram of the nanoparticle tagging of proteins anchored to gold nanostructure surfaces.

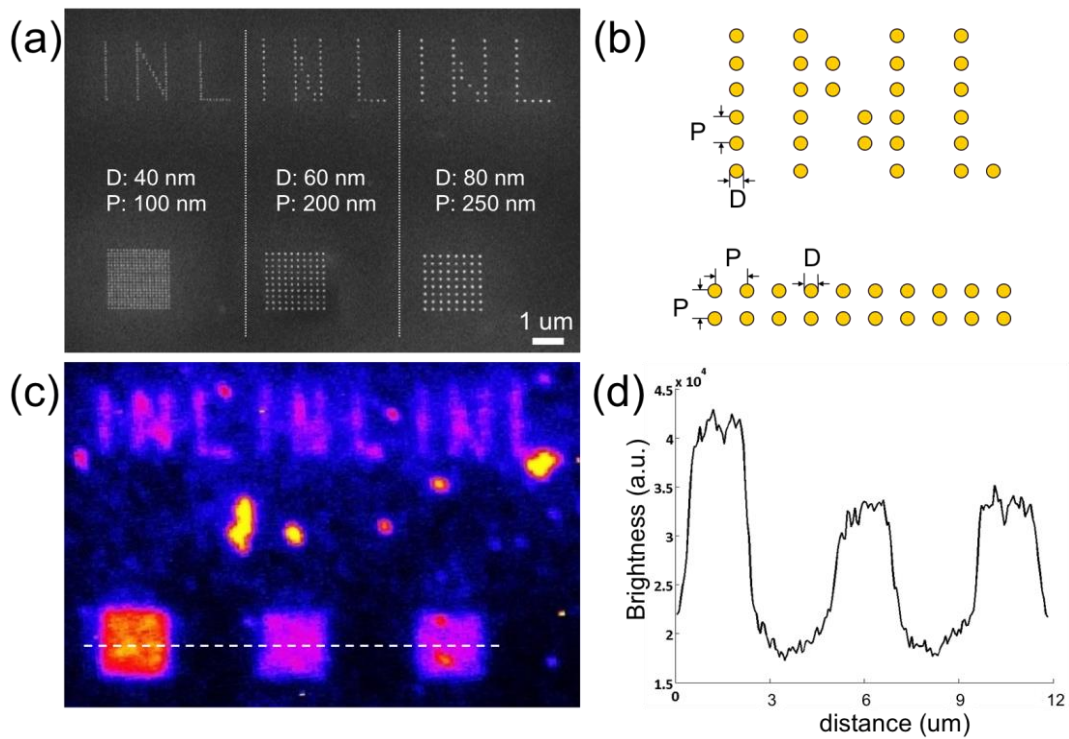

**Figure 2.** Localization of cortexillin I$_{272-352}$ around patterned gold nanostructures. (a) Scanning electron microscopy of the gold nanodot arrays with various diameters (D) and pitches (P). (b) Diagram showing the arrangement of the gold-nanodot patterns shown in (a). (c) Widefield fluorescence microscopy image of the gold-nanodot patterns incubated with cortexillin I$_{272-352}$, primary antibody against FLAG, and FITC-tagged secondary antibody. (d) Fluorescence intensity profile along the white dotted line in (c).

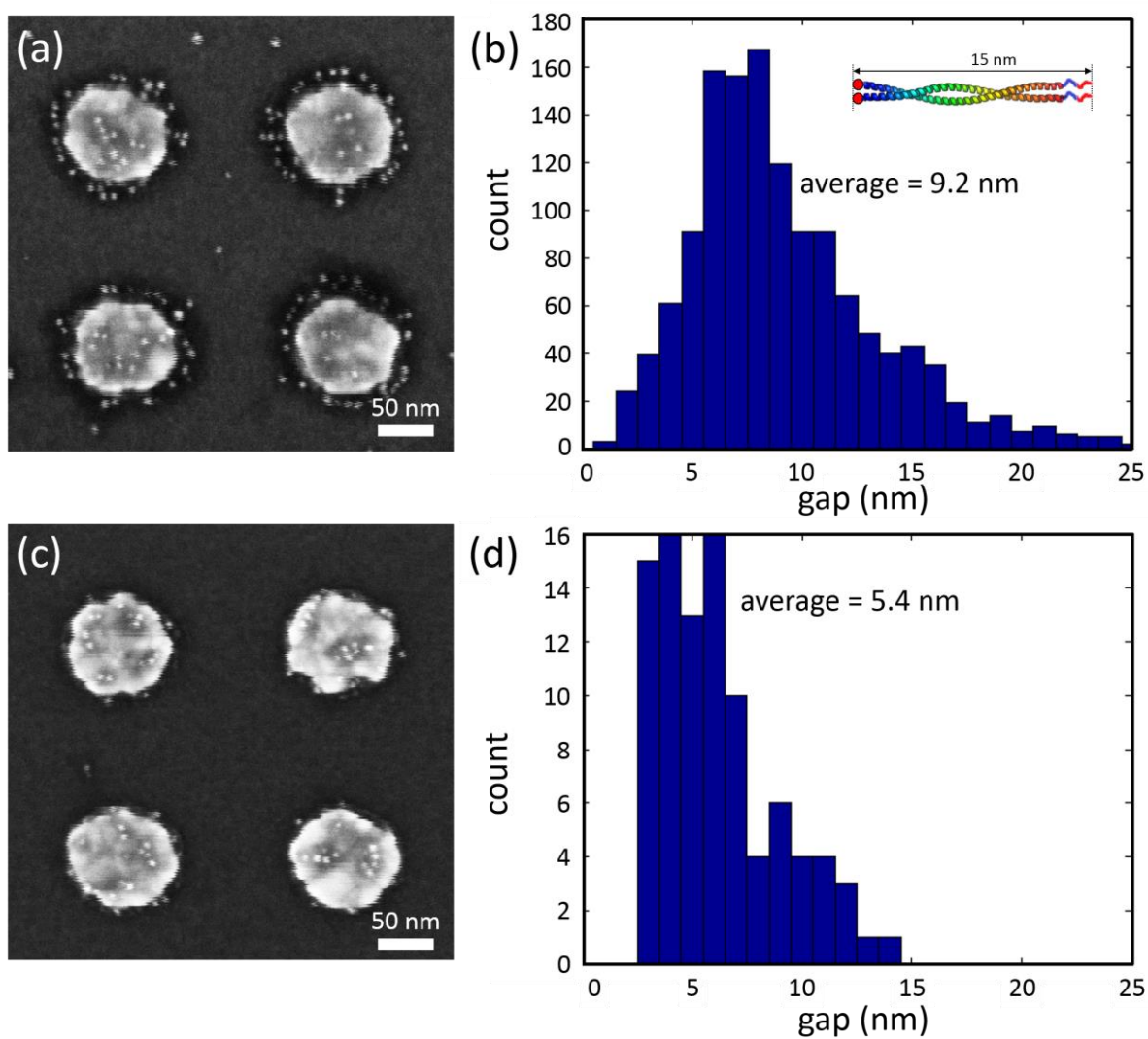

**Figure 3**. Cortexillin I$_{272-352}$ attached to the gold nanodots at room temperature (20°C) vs. 80°C. (a,c) SEM images of the gold nanodots incubated with cortexillin I$_{272-352}$ and gold nanoparticles at (a) room temperature and (c) 80°C. (b,d) Histograms showing the distributions of the gaps between the gold nanodot surface and nanoparticles attached to the N-terminus of cortexillin I$_{272-352}$ incubated at (b) room temperature and (d) 80°C. The number of posts analyzed was 135 and 71 for (b) and (d), respectively. Total number of gold nanoparticles was 1414 and 95 for (b) and (d), respectively.

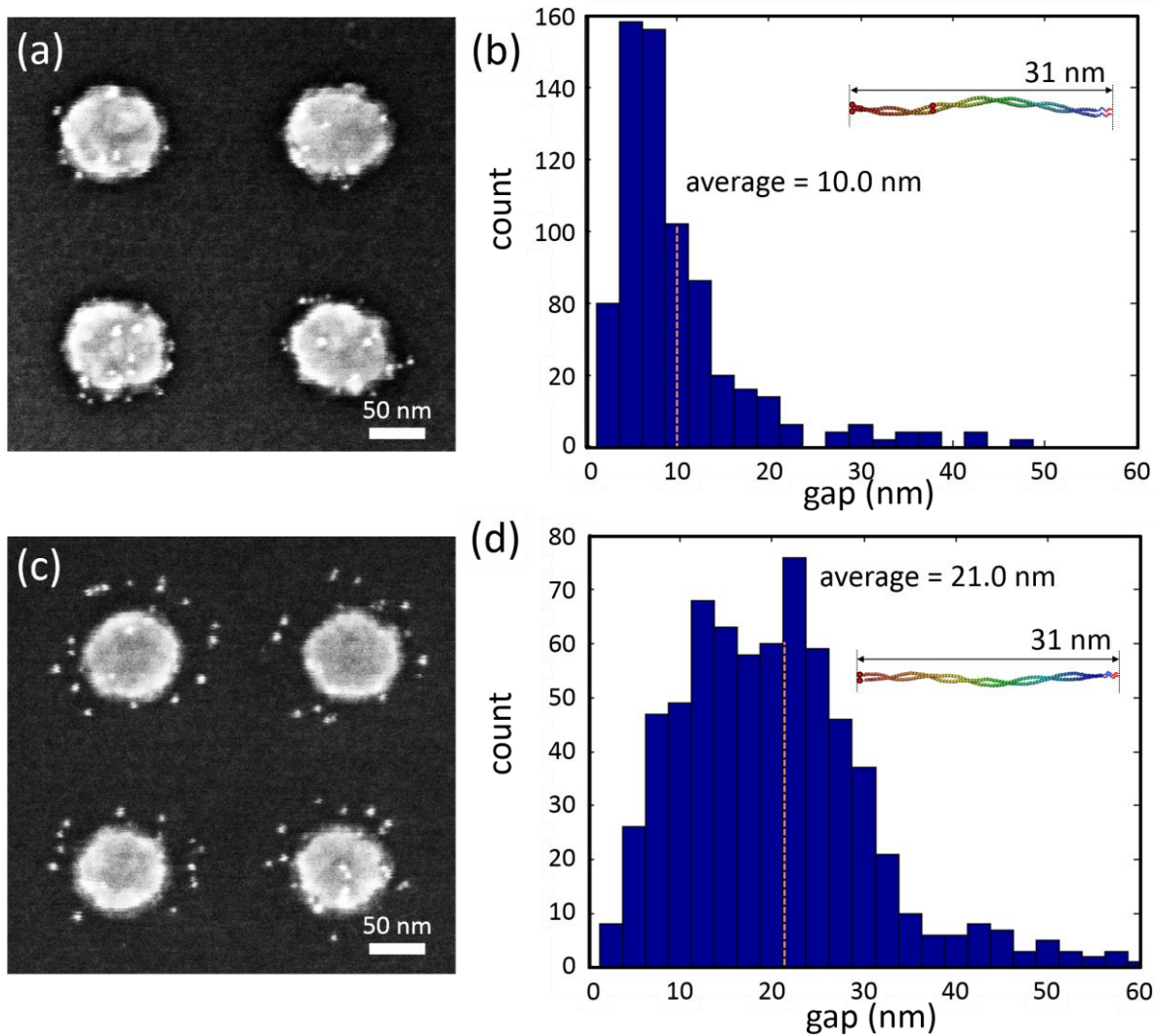

**Figure 4.** Tropomyosin variants templated by gold nanodots. (a,c) SEM images of gold nanodots incubated with (a) native tropomyosin (with two cysteine residues) and gold nanoparticles or (c) tropomyosin C190A and gold nanoparticles. (b,d) Histograms showing the distributions of the gap sizes between the gold nanodots and (b) native tropomyosin or (d) C190A tropomyosin. The number of posts analyzed was 63 and 51 for (b) and (d), respectively. Total number of gold nanoparticles was 232 and 680 for (b) and (d), respectively.

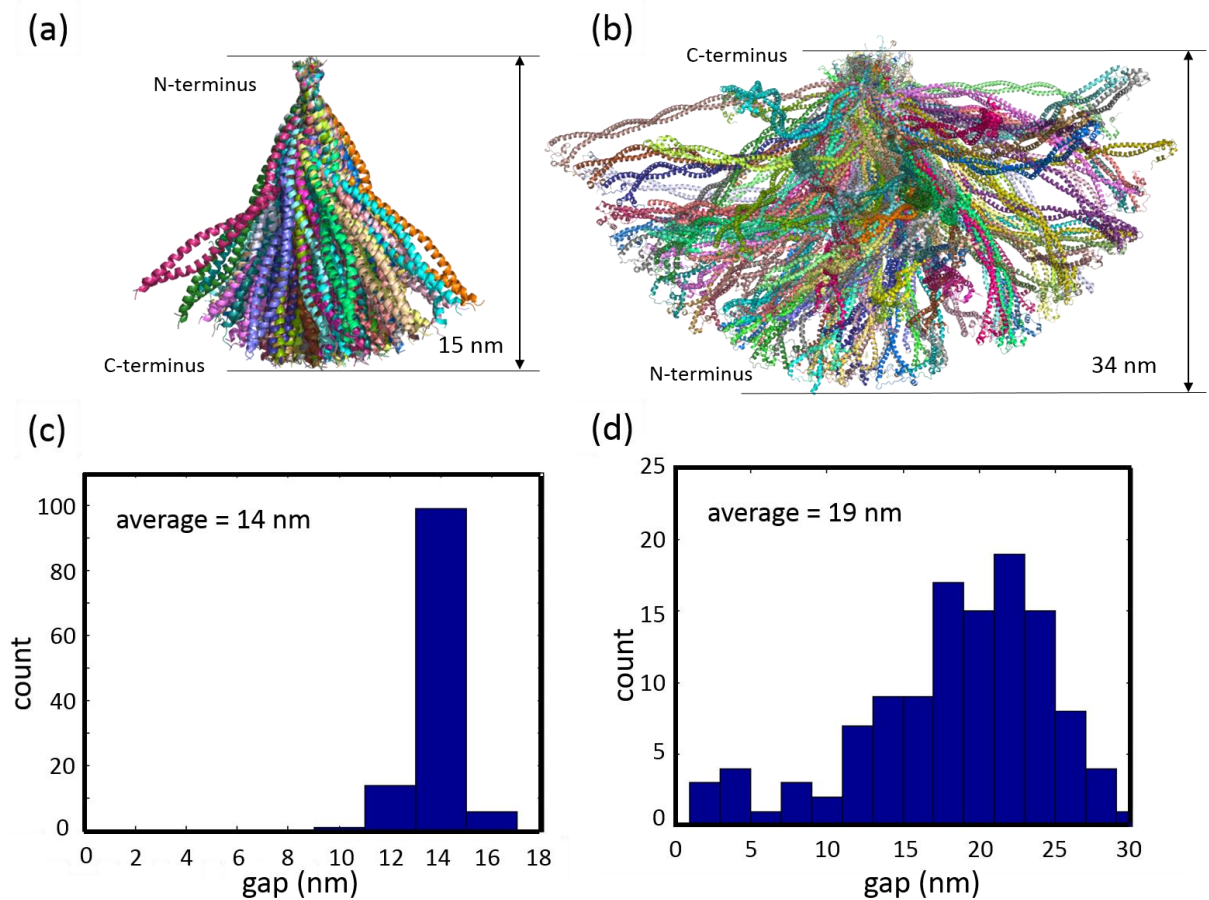

**Figure 5**. Molecular dynamics simulation of cortexillin I$_{272-352}$ and tropomyosin. (a) Side view of the simulated cortexillin I$_{272-352}$. (b) Side view of the simulated tropomyosin. (c,d) Histogram showing the distribution of the distances between the two termini of (c) cortexillin I$_{272-352}$ (d) tropomyosin, if (c) the N-terminus of cortexillin I$_{272-352}$ (d) the C-terminus of tropomyosin was attached to the gold nanodot (n=120).

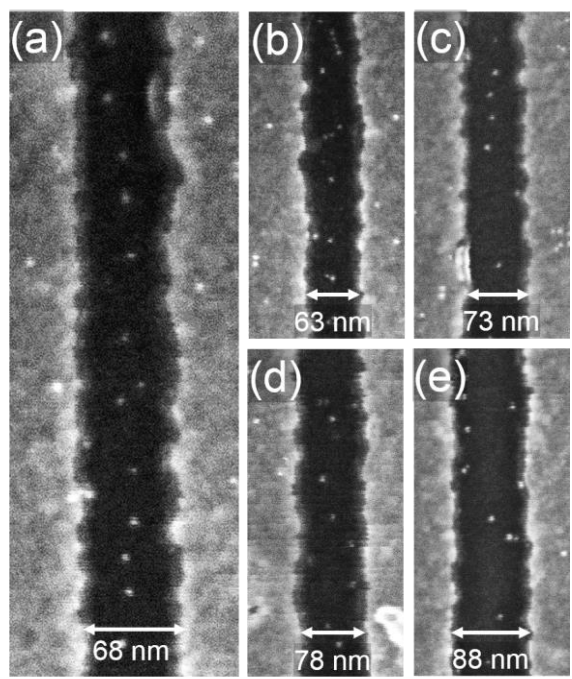

**Figure 6**. Scanning electron microscopy of gold nanoparticles templated by tropomyosin C190A and a gold trench. The width of the trench is (a) 68 nm; (b) 63 nm; (c) 73 nm; (d) 78 nm; and (e) 88 nm.

# Supporting Information

**Title: Single-molecule study on the orientation of coiled-coil proteins attached to gold nanostructures**

*Jae-Byum Chang[+], Yong Ho Kim[+], Evan Thompson, Young Hyun No, Nam Hyeong Kim, Jose Arrieta, Vitor R. Manfrinato, Amy E. Keating\*, Karl K. Berggren\* ([+]equally contributed)*

**1. End-to-end distance of cortexillin and tropomyosin**

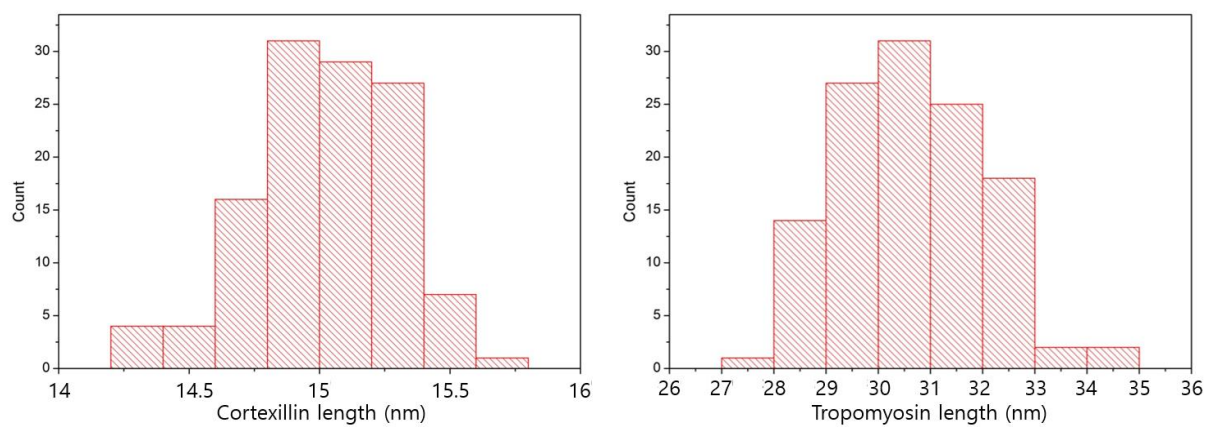

**Figure S1**. End-to-end distance distribution of cortexillin and tropomyosin measured by the MD simulations of cortexillin (left) and tropomyosin (right). The averages were 15 nm (left) and 31 nm (right). The standard deviations were 0.3 nm (left) and 1.4 nm (right).

## 2. Statistical analysis of the gap size of the gold nanoparticles templated by a gold trench and C190A tropomyosin.

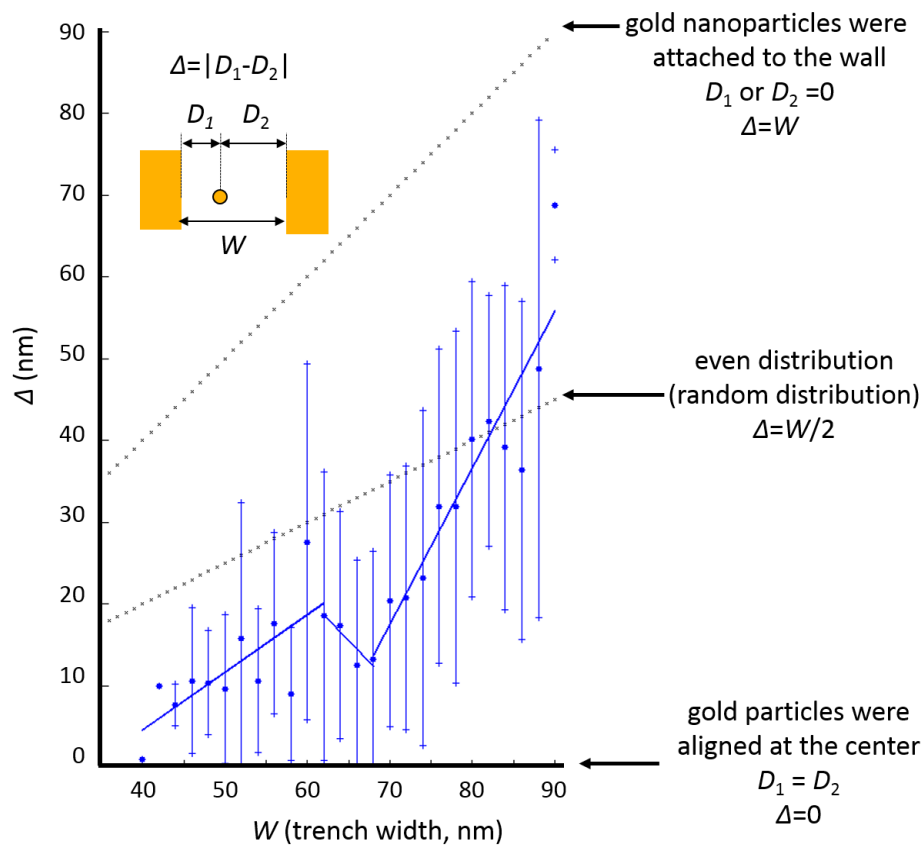

**Figure S2**. Distribution of the differences in distances to each wall of coiled coil-attached nanoparticles inside the gold trench. To understand the distribution of gold nanoparticles inside the trench, the distances between one gold particle and two side walls ($D_1$ and $D_2$) were measured. If a gold nanoparticle was positioned at the center of the trench, $D_1$ should be equal to $D_2$; in other words, $|D_1-D_2|$ should be zero. If a gold nanoparticle was attached to one of the two sidewalls, $|D_1-D_2|$ should be equal to the width of the trench. If a gold nanoparticle was randomly distributed inside the trench, the average of $|D_1-D_2|$ over many gold nanoparticles should be $W/2$ (i.e., half the width of the trench). In the graph depicted here, the *x*-axis represents the width of the trench and the *y*-axis represents the difference between $D_1$ and $D_2$ ($\Delta=|D_1-D_2|$). When the width of the trench was between 60 nm and 70 nm, $\Delta$ decreased and approached to the *x*-axis ($\Delta=0$), indicating that the tendency of the gold nanoparticles to be aligned at the center of the trench increased. $\varDelta$ was at a minimum when the width of the trench was 68 nm. When the width of the trench was greater than 70 nm, $\varDelta$ increased rapidly, indicating that the tendency of gold nanoparticles to be aligned at the center decreased and, instead, they tended to be randomly scattered inside the trench. The blue lines show ±1 standard deviation. The number of analyzed trenches was 5, and the number of total gold nanoparticles was 363. Figure S3 shows the SEM images of the 5 trenches used for this statistical analysis.

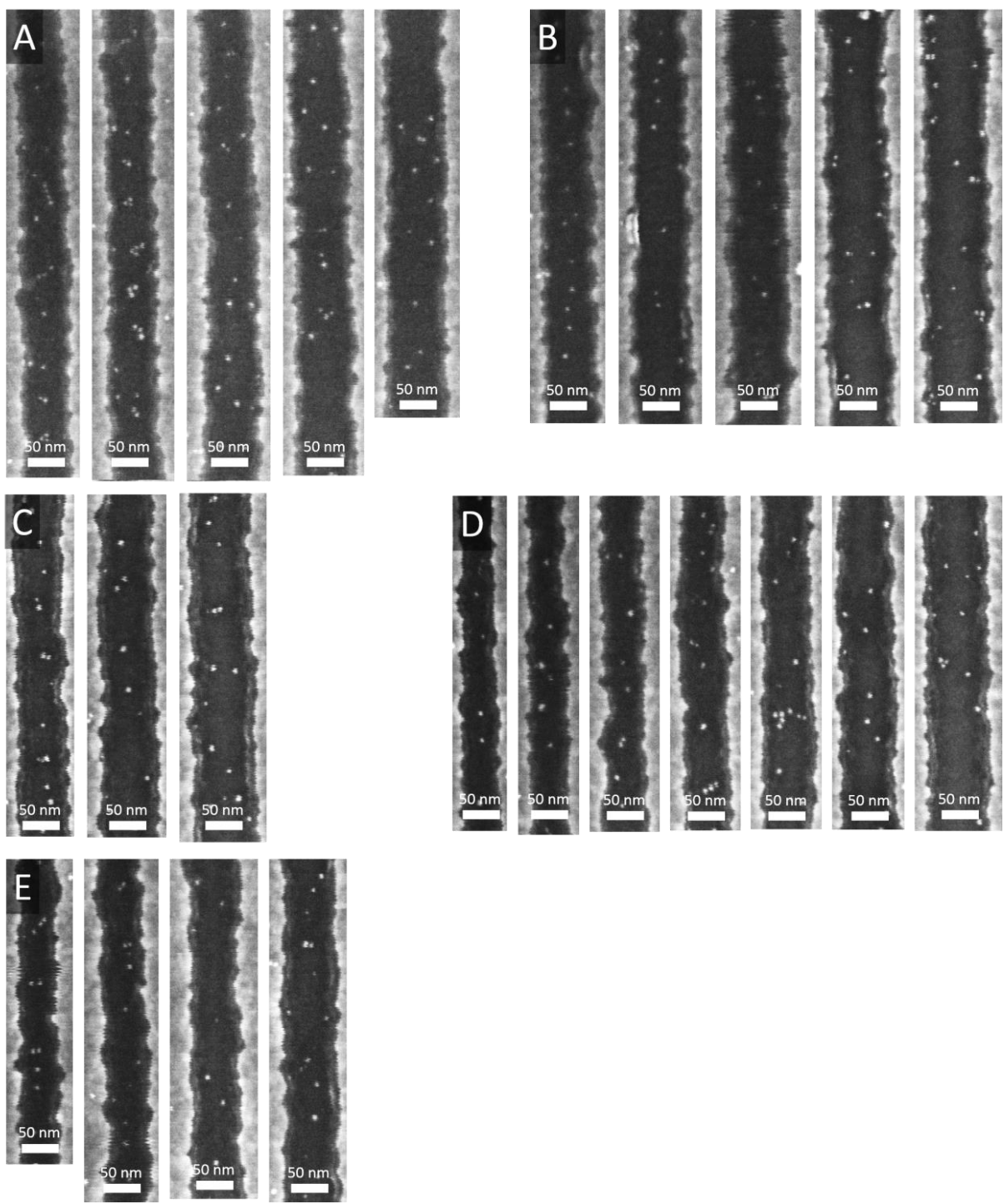

**Figure S3.** SEM images of five trenches that were analyzed in Figure S3. (A-E) Each panel shows SEM images a single long trench, with a width varying from 40 nm to 90 nm. The width of the trenches increases from left to right.

## 3. Fabrication process

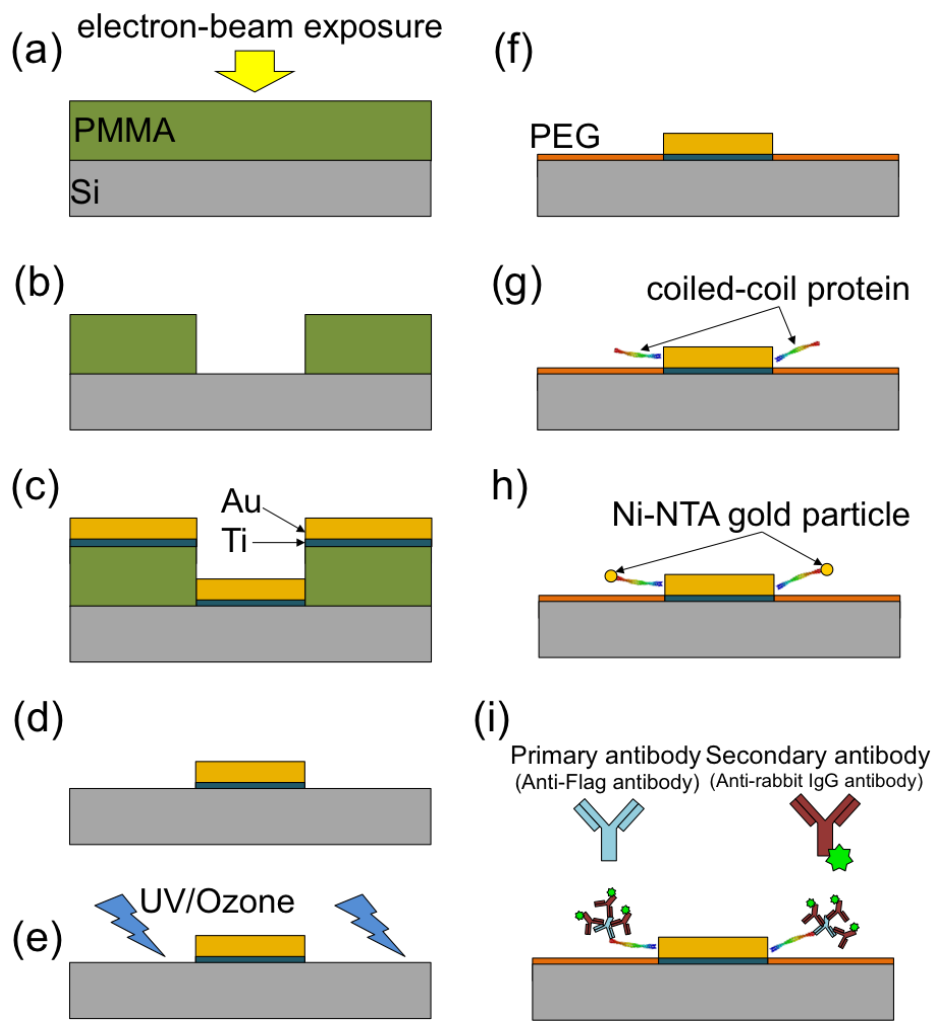

**Figure S4**. Gold nanodot fabrication process and two methods of visualizing the arrangements of the attached proteins around the gold nanodots.

## 4. Protein sequences

Protein sequence of native tropomyosin

```
M HHHHHH DYKDDDDK GG
           10         20         30         40         50         60
MDAIKKKMQM LKLDKENALD RAEQAEADKK AAEDRSKQLE DELVSLQKKL KGTEDELDKY
           70         80         90        100        110        120
SEALKDAQEK LELAEKKATD AEADVASLNR RIQLVEEELD RAQERLATAL QKLEEAEKAA
          130        140        150        160        170        180
DESERGMKVI ESRAQKDEEK MEIQEIQLKE AKHIAEDADR KYEEVARKLV IIESDLERAE
          190        200        210        220        230        240
ERAELSEGKC AELEEELKTV TNNLKSLEAQ AEKYSQKEDR TEEEIKVLSD KLKEAETRAE
          250        260        270        280  284
FAERSVTKLE KSIDDLEDEL YAQKLKYKAI SEELDHALND MTSI    G C
```

DNA sequence of native tropomyosin

```
     atgcatcatc accatcacca cgattacaag gatgacgacg ataagggtgg t
  1  atggacgcca tcaagaagaa gatgcagatg ctgaagctcg acaaggagaa cgccttggat
 61  cgagctgagc aggcggaggc cgacaagaag gcggcggaag acaggagcaa gcagctggaa
121  gatgagctgg tgtcactgca aaagaaactc aagggcaccg aagatgaact ggacaaatac
181  tctgaggctc tcaaagatgc ccaggagaag ctggagctgg cagagaaaaa ggccaccgat
241  gctgaagccg acgtagcttc tctgaacaga cgcatccagc tggttgagga agagttggat
301  cgtgcccagg agcgtctggc aacagctttg cagaagctgg aggaagctga aaggcagca
361  gatgagagtg agagaggcat gaaagtcatt gagagtcgag cccaaaaaga tgaagaaaaa
421  atggaaattc aggagatcca actgaaagag gccaagcaca ttgctgaaga tgccgaccgc
481  aaatatgaag aggtggcccg taagctggtc atcattgaga gcgacctgga acgtgcagag
541  gagcgggctg agctctcaga aggcaaatgt gccgagcttg aagaagaatt gaaaactgtg
601  acgaacaact tgaagtcact ggaggctcag gctgagaagt actcgcagaa ggaagacaga
661  tatgaggaag agatcaaggt cctttccgac aagctgaagg aggctgagac tcgggctgag
721  tttgcggaga ggtcagtaac taaattggag aaaagcattg atgacttaga agacgagctg
```

```
781 tacgctcaga aactgaagta caaagccatc agcgaggagc tggaccacgc tctcaacgat
841 atgacttcca ta
    ggt tgt taa
```

Protein sequence of tropomyosin (C190A)

```
M HHHHHH DYKDDDDK GG
          10         20         30         40         50         60
MDAIKKKMQM LKLDKENALD RAEQAEADKK AAEDRSKQLE DELVSLQKKL KGTEDELDKY
          70         80         90        100        110        120
SEALKDAQEK LELAEKKATD AEADVASLNR RIQLVEEELD RAQERLATAL QKLEEAEKAA
         130        140        150        160        170        180
DESERGMKVI ESRAQKDEEK MEIQEIQLKE AKHIAEDADR KYEEVARKLV IIESDLERAE
         190        200        210        220        230        240
ERAELSEGKA AELEEELKTV TNNLKSLEAQ AEKYSQKEDR TEEEIKVLSD KLKEAETRAE
         250        260        270        280   284
FAERSVTKLE KSIDDLEDEL YAQKLKYKAI SEELDHALND MTSI    G C
```

DNA sequence of tropomyosin(C190A)

```
    atgcatcatc accatcacca cgattacaag gatgacgacg ataagggtgg t
  1 atggacgcca tcaagaagaa gatgcagatg ctgaagctcg acaaggagaa cgccttggat
 61 cgagctgagc aggcggaggc cgacaagaag gcggcggaag acaggagcaa gcagctggaa
121 gatgagctgg tgtcactgca aaagaaactc aagggcaccg aagatgaact ggacaaatac
181 tctgaggctc tcaaagatgc ccaggagaag ctggagctgg cagagaaaaa ggccaccgat
241 gctgaagccg acgtagcttc tctgaacaga cgcatccagc tggttgagga agagttggat
301 cgtgcccagg agcgtctggc aacagctttg cagaagctgg aggaagctga aaggcagca
361 gatgagagtg agagaggcat gaaagtcatt gagagtcgag cccaaaaaga tgaagaaaaa
421 atggaaattc aggagatcca actgaaagag gccaagcaca ttgctgaaga tgccgaccgc
481 aaatatgaag aggtggcccg taagctggtc atcattgaga gcgacctgga acgtgcagag
541 gagcgggctg agctctcaga aggcaaagcg gccgagcttg aagaagaatt gaaaactgtg
```

```
601 acgaacaact tgaagtcact ggaggctcag gctgagaagt actcgcagaa ggaagacaga
661 tatgaggaag agatcaaggt cctttccgac aagctgaagg aggctgagac tcgggctgag
721 tttgcggaga ggtcagtaac taaattggag aaaagcattg atgacttaga agacgagctg
781 tacgctcaga aactgaagta caaagccatc agcgaggagc tggaccacgc tctcaacgat
841 atgacttcca ta
    ggt tgt taa
```

## Protein Sequence of Cortexillin

```
M SYKCGGS
         10         20         30         40         50         60
AYRAKEEKAR LESSKNEMAN RLAGLENSLE SEKVSREQLI KQKDQLNSLL ASLESEGAER
         70         80         90        100        110        120
EKRLRELEAK LDETLKNLEL EKLARMELEA RLAKTEKDRA ILELKLAEAI DEKSKLEQQI
    126
EATRIR
EFG DYKDDDDKG HHHHHH
```

## DNA sequence of Cortexillin

```
      atgtcttaca aatgcggtgg atcc
  1 gcatatcgtg cgaaagaaga aaaagcgcgt ctggaaagct ctaaaaacga atggcgaac
 61 cgtctggcgg gtctggagaa ctccctcgaa tccgagaaag tttctcgtga acagctgatc
121 aaacagaaag accagctcaa ctctctgctg gcatccctgg aatctgaagg tgcggaacgc
181 gaaaaacgtc tgcgtgaact cgaagcgaaa ctcgatgaaa ccctgaaaaa cctggagctg
241 aaggatagag ctatcttgga attgaaatta gctgaagcca tcgatgaaaa atcaaaactc
301 atcctggaac tgaaactggc ggaagcgatc gacgaaaaat ctaagctcga acagcagatt
361 gaggcaactc gtattcgt
    gaattcggc gactacaaagacgatgacgacaaaggc caccatcatcaccaccac taa
```

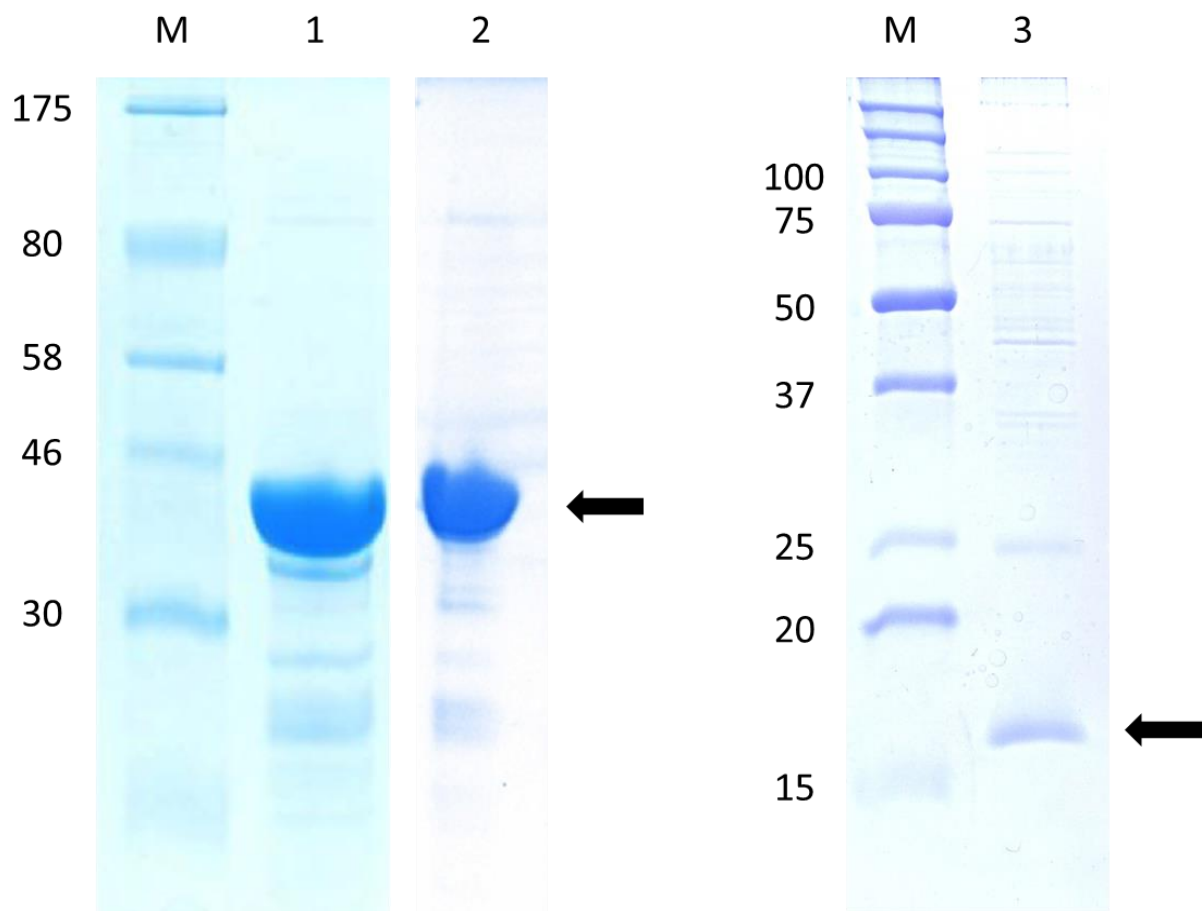

**Figure S6**. SDS-PAGE. Lane M; molecular mass markers (kDa), Lane 1; Tropomyosin Mutant C190A, Lane 2; Native Tropomyosin, Lane 3; Cortexillin.